\documentclass[aoas,nameyear,dvips]{arximspdf}
\usepackage{dcolumn}
\usepackage{graphicx}

\doi{10.1214/10-AOAS334}
\volume{4}
\issue{3}
\pubyear{2010}
\firstpage{1256}
\lastpage{1271}

\makeatletter
\newcolumntype{d}[1]{D{.}{.}{#1}}
\newtheorem{theorem}{Theorem}
\newproclaim{remark}{Remark}
\newproclaim{algorithm}{Algorithm}
\newcommand{\eqref}[1]{(\ref{#1})}
\renewcommand{\citep}[1]{\citeauthor{#1} \citeyear{#1}}
\makeatother

\begin{document}
\begin{frontmatter}

\title{Detection of radioactive material entering national ports: A
Bayesian approach\\ to radiation portal data\thanksref{T1}}
\thankstext{T1}{Supported in part by NSF Grant CBET-0736134.}
\runtitle{A Bayesian approach to radiation portal data}

\begin{aug}
\author{\fnms{Siddhartha R.} \snm{Dalal}\ead[label=e1]{sdalal@rand.org}\corref{}}
\and
\author{\fnms{Bing} \snm{Han}\ead[label=e2]{bhan@rand.org}}
\runauthor{S. R. Dalal and B. Han}
\affiliation{RAND Corporation}
\address{RAND Corporation\\
1776 Main St.\\
Santa Monica, California 90401\\
USA\\
\printead{e1}} 
\end{aug}

\received{\smonth{8} \syear{2009}}
\revised{\smonth{1} \syear{2010}}

%
\begin{abstract}
Given the potential for illicit nuclear material being used for
terrorism, most ports now inspect a large number of goods entering
national borders for radioactive cargo. The U.S. Department of Homeland
Security is moving toward one hundred percent inspection of all
containers entering the U.S. at various ports of entry for nuclear
material. We propose a Bayesian classification approach for the
real-time data collected by the inline Polyvinyl Toluene radiation
portal monitors. We study the computational and asymptotic properties
of the proposed method and demonstrate its efficacy in simulations.
Given data available to the authorities, it should be feasible to
implement this approach in practice.
\end{abstract}

%

\end{frontmatter}

\section{Introduction}\label{sec1}
With increased terrorism around the world and instability in some
nuclear-capable nations, there is a growing national safety concern
about terrorists bringing illicit nuclear materials into the U.S.
Substantial efforts have gone into devising strategies for inspecting
containers and intercepting various types of illicit nuclear material.
In the U.S. there are 307 ports of entry representing 621 official air,
sea and land border crossing sites, through which approximately 57,000
containers enter the borders every day. For effective inspection
without increasing traffic congestion, the U.S. Department of Homeland
Security has adopted a multilayered approach to inspection, which
consists of an analysis of customs documents, followed by an inline
automatic inspection of the containers, and an offline stringent manual
inspection for suspicious containers. For more details of the process
and the corresponding risk analysis, we refer to \citet{Wein2006} and
\citet{Martonosi2006}.

One objective of the inline preliminary inspection procedure is to
identify radioactive cargo that is being shipped in a container.
The inline preliminary inspection procedure consists of scanning
containers in a radiation portal monitor (RPM) via a gamma ray
Polyvinyl Toluene (PVT) scanner. Currently, 98\% of incoming containers
go through this radiation scanning situated at most ports of entry.
Based on the data collected during scanning, the inline inspection
procedure makes a quick automatic decision to let a container pass or
to scrutinize it further with the offline process. In addition, a few
containers are randomly selected for offline inspection. For example,
the Los Angeles Times (11/26/2004) reported that at the Los Angeles
port in 2004, around 12,000 containers arrived daily and, on average,
43 were inspected by hand.

The design of the RPM consists of a drive-through portal with passive
PVT sensors that detect gamma rays emitting from a source. At an
inspection point, the container is driven through the portal at low
speed (4--5 mph), taking around 20 seconds. The portal captures the
radiation counts at every 0.1-second interval in a number of energy
channels ranging from low to high. For example, the portals
manufactured by SAIC in certain configurations have 256 channels. The
next generation of portals based on Sodium Iodide Scintillators will
have 1024 channels. The collected data consist of the radiation count
in each of the channels at every 0.1-second interval, accumulated over 1
second. In practice, the data are further aggregated over multiple
channels in a number of nonoverlapping coarser windows. Typical
configurations involve 2--8 nonoverlapping exhaustive windows from
very low energy to very high energy.
In our paper we focus on nonoverlapping energy windows, referred to as
windows in the rest of this paper. For an excellent exposition of the
details related to energy windowing, we refer to \citet{Ely2004} and
\citet{Ely2006}.

\begin{figure}

\includegraphics{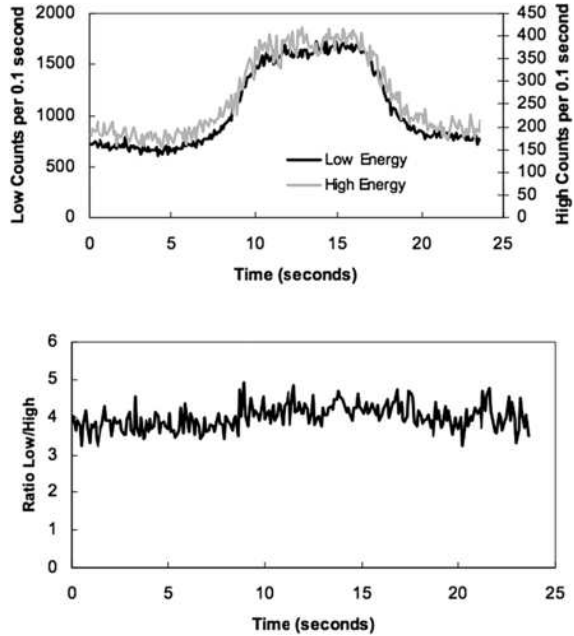}

\caption{Gamma counts as a container of NORM passes a 2-window PVT. The
container passes through the portal for around 23 seconds.
The upper frame shows the superposition of the low- and high-energy
windows with different scales. The lower frame gives the ratio of
counts between the two windows.}\label{fig:nex}
\end{figure}

Given that the distance from the source changes as a container is
driven through the portal, radiation counts will also change. The upper
frame of Figure \ref{fig:nex} shows an example of radiation count data
as a container is rolling through a radiation portal (courtesy of
Pacific Northwest National Laboratory).
The figure depicts readings of a container passing through a 2-window
system corresponding to the high- and low- energy windows. The upper
frame of the figure superposes the readings of~the two windows with the
different scales on the left- and right-hand axes, while the~lower
frame plots the ratio of low- to high-energy readings.
Currently, only the total count corresponding to the distance which
yields the maximal total count is used for detection purposes. When the
total count exceeds a preset threshold, a~container is classified as
potentially dangerous. However, this crude method may fail to detect
dangerous man-made sources in small quantities when mixed with
naturally occurring radioactive materials. In Section \ref{sec5} we
show a
numerical example where all containers have approximately the same mean
total count, which the current inspection cannot differentiate.

In this paper we propose a decision theoretic approach based on
Bayesian methods for identifying suspicious containers using the data
collected during the inline inspection. This paper is organized as
follows. In Section \ref{sec2} we introduce the available RPM data
collected by
the inline inspection system, and describe a statistical model for the
RPM data based on a Poisson process. In Section \ref{sec3} we introduce the
machinery needed for a Bayesian approach and propose a naive Bayesian
classifier. In Section \ref{sec4} we show that the proposed procedure is
asymptotically accurate in the sense that the probability of
misclassification goes to zero as the mean radiation counts become
large. In Section \ref{sec5} we explore the efficacy of our procedure by
numerical examples based on real energy spectra for exact
classification. We also propose an algorithm to simplify the classifier
when the objective is limited to discriminating between dangerous and
nondangerous materials. Finally, this paper concludes with a discussion
of some of the implementation challenges and future extensions.

\section{The Poisson model for radiation emission data}\label{sec2}
A container may include man-made radioactive material of high concern,
including Highly Enriched Uranium (HEU) or Weapons Grade Plutonium
(WGPu), as well as other common classes of cargo that have been
officially declared in filings with Customs.
Some common classes are naturally occurring radioactive materials
(NORM) and are often misclassified based on the currently deployed
methodologies during inline inspection. Some common NORM classes
include fertilizer, kitty litter and refractory material. The
radioactivity in NORM is caused by some ingredients with natural
radioactivity, for instance, clay from some regions in Mexico.

For the remainder of this paper, we tackle the objective of classifying
a container into one of $K$ classes.
Initially we consider each class to be either a NORM or a man-made
nuclear material. Later on we shall discuss the situation involving
mixtures of materials.
We first introduce the notation for the radiation data. Let~$Y$ denote
the class variable, $Y=k$ if the container being inspected is in the
$k$th class. Let $\mathbf{Z}$ denote all of the radiation count data
obtained at a RPM with $B$ windows corresponding to a particular
container. Let $\mathbf{R}_d=(R_{1,d},\ldots,R_{B,d})'$ denote the
vector of counts at a distance $d=d(t)$ away from the detectors. Time
$t$ is used as the surrogate of $d$ in Figure \ref{fig:nex}. $\mathbf
{Z}$ is a $B \times T$ matrix
%
\begin{equation}
\mathbf{Z} = \bigl(\mathbf{R}_{d(1)},\mathbf{R}_{d(2)}, \ldots, \mathbf
{R}_{d(T)}\bigr) =
\pmatrix{R_{1,d(1)} & R_{1,d(2)} & \ldots& R_{1,d(T)} \cr
R_{2,d(1)} & R_{1,d(2)} & \ldots& R_{2,d(T)} \cr
\ldots & \ldots & \ldots& \ldots\cr
R_{B,d(1)} & R_{B,d(2)} & \ldots& R_{B,d(T)}}
,
\end{equation}
where $T$ is the total number of sampling times, $t=1,\ldots, T$.
Finally, let $N_d$ denote the total counts at distance $d$, $N_d = \sum
_b R_{b,d}=\mathbf{1'R}_d$.

Poisson models have been frequently used in modeling radiation counts
and other types of emission counts [\cite{Karlin1998}].
We use the following Poisson process model \eqref{eqn:model} for the
RPM count data, which will be subsequently used to build a naive
Bayesian classifier in the next section. Let $M$ be the quantity of
material in a container, and let $\lambda_{b,k}$ be the emission rate
for a unit quantity of class $k$ in window $b$ at the distance $d=0$.
Given that a container has $M$ quantity of the $k$th material class and
is at distance $d$ from the detector,
%
\begin{eqnarray}\label{eqn:model}
&&R_{b,d} | (Y=k, d, M) \sim\operatorname{Poisson}(g(d,M)\lambda
_{b,k})\\
\eqntext{b=1,\ldots,
B, d=d(1),\ldots, d(T),}
\end{eqnarray}
where all $R_{b,d}$ are independent, and $g(d,M)$ is an unknown
function. One would expect $g(d,M)$ to be decreasing in distance $d$
and increasing in quantity $M$, with $g(0,1)=1$. The specific form of
$g$ depends on the physical mechanism of sensors, signal processing and
various environmental factors.

\citet{Ely2004} and \citet{Ely2006} delve into a substantive
discussion of $g(d,M)$ through the amount of information available in
$N_d$, which they find is very unreliable. Moreover,
the function $g$ is associated not only with distance $d$ and amount of
material $M$, but also with various environmental conditions (e.g.,
weather and background noise), vehicle shielding (which varies from
container to container), and a container's angular placement relative
to the detectors (which changes as the vehicle traverses through the
portal), and sensitivity of sensors. Thus, $g$ is a function of many
more variables than just the quantity and distance. Modeling $g$ would
be very difficult since many factors are unknown or not collected. We
will not pursue this parametric line of inquiry here because of the
inherent difficulties mentioned above and the lack of publicly
available data.

Given the difficulties in modeling $g$, we consider $g$ as a nuisance
parameter and remove it for inference by appropriate conditioning. To
realize this, note that $N_d$ follows independent Poisson distributions
with mean $g(d,M)\sum_b \lambda_{b,k}$. By construction, the
conditional distribution is
\begin{eqnarray*}
P(\mathbf{R}_d|N_d,Y=k,d,M) &\propto& P(R_{1,d},\ldots,R_{B,d},N_d |
Y=k,d,M) \\
&=& P(R_{1,d},\ldots,R_{B,d}| Y=k,d,M).
\end{eqnarray*}
We then have
%
\begin{eqnarray}
&&P(R_{1,d},\ldots,R_{B,d} | Y=k,d,M,N_d) \nonumber\\
&&\qquad\propto\prod_b [g(d,M)
\lambda_{b,k}]^{R_{b,d}}/(R_{b,d}!)\\
&&\qquad\propto \frac{N_d!}{R_{1,d}!\cdots R_{B,d}!} \biggl(\frac{\lambda
_{b,k}}{\lambda_k}\biggr)^{R_{b,d}}
\times\frac{[g(d,M)\lambda_k]^{N_d}}{N_d!},\nonumber
\end{eqnarray}
where $\lambda_k=\sum_b \lambda_{b,k}$. The marginal distribution of
$N_d$ is easily shown to be $\operatorname{Poisson}(g(d,M)\lambda_k)$. It follows that
the conditional distribution is
%
\begin{equation}\label{eqn:cond}
\mathbf{R}_d|(Y=k,d,M,N_d) \sim \operatorname{multinomial}(N_d; p_{1,k},\ldots,p_{B,k}),
\end{equation}
where $p_{b,k} = \lambda_{b,k}/\lambda_k, b=1,\ldots, B$. Note that $0
\le p_{b,k} \le1$ and $\sum_b p_{b,k}=1$.
Hence, given $N_d$, $\mathbf{R}_d$ follows a multinomial distribution
with the probability parameters $p_{1,k},\ldots,p_{B,k}$ independent of
$d$. In addition, by model \eqref{eqn:model}, $\mathbf{R}_{d(t)},
t=1,\ldots, T$, at different distances are all independent. It follows
that, if we were to base our inference conditioned on $N_d$, that is,
the total radiation count at time $d$ for all windows, we would not
require any specific functional model for $g(d,M)$ and~$N_d$. The lower
frame of Figure \ref{fig:nex} corresponds to a real data example that
validates this assumption. It can be seen that the ratio in counts
between the two windows is approximately constant during the actual
scan lasting around 23 seconds.

Let $\mathbf{p}_k = (p_{1,k},\ldots, p_{B,k})'$ be named as the \textit
{energy spectrum} of the $k$th class in the rest of this paper.
The energy spectrum is unique for a material class and invariant to $g(d,M)$.
By the above derivation, we can simply focus on the energy spectrum
$\mathbf{p}_k$. This is more pertinent and stable than the total
radiation count $N_d$, which depends on the unknown $g(d,M)$. The
disadvantage of using the conditioning likelihood is that it would not
be possible to distinguish between two material classes that have an
identical energy spectrum. We also remark that \citet{Ely2006}
suggested that an energy windowing method should be used in conjunction
with a gross counting threshold set at a relatively insensitive level.

The proposed conditional multinomial model has additional advantages as
well. First, we can aggregate or disaggregate windows while keeping the
new counts multinomial. Further, since our multinomial distribution
depends on $d$ only through $N_d$, we can pool the information from the
counts obtained at many distances. We do not have to depend upon the
radiation data at the maximum of $N_d$ as the current approach does.
Finally, conditioning on $N_d$ allows us to implement a classification
approach parallel to the naive Bayes classifier. We describe the
development of the Bayesian classifier in the next two sections.

\section{Bayesian classifier}\label{sec3}
For an introduction to the naive Bayes classification, we refer to
\citet{Ye2003} and \citet{Klosgen2002}. Here we construct a
corresponding generative naive Bayesian classifier. We start with two
distinct material classes $k$ and $k'$. By the Bayes theorem,
%
\begin{equation}\label{eqn:init}
\frac{P(Y=k|\mathbf{Z})}{P(Y=k'|\mathbf{Z})}=\frac{P(Y=k)P(\mathbf
{Z}|Y=k)}{P(Y=k')P(\mathbf{Z}|Y=k')},
\end{equation}
where $P(Y=k)$ and $P(Y=k')$ are the prior probability. Apparently, a
container is classified as class $k$ if \eqref{eqn:init} is larger than
1. Generalizing this to $K$ classes, we have the following classifier:
%
\begin{equation}\label{eqn:rule}
Y = \arg\max_{1 \le k \le K} [P(Y=k)P(\mathbf{Z}|Y=k)].
\end{equation}
It is easily shown [\citet{Ferguson1967}] that this rule is optimal in
terms of Bayes
risk as long as the cost of misclassification is the same across the categories.

The naive Bayesian classifier \eqref{eqn:rule} has a convenient form
for computation. Further conditioning on all $N_{d(t)}$, $t=1,\ldots, T$,
by \eqref{eqn:cond}, we have
%
\begin{equation}
P\bigl(\mathbf{Z}|Y=k,N_{d(1)},\ldots,N_{d(T)}\bigr) \propto\prod_{b}
p_{b,c}^{\sum_d R_{b,d}}.
\end{equation}
Equivalently, the classifier can be represented by the log likelihood
%
\begin{eqnarray}\label{eqn:finalform}
Y &= &\arg\max_{1 \le k \le K} \biggl[\sum_{b,d} R_{b,d} \log p_{b,k} + \log
P(Y=k)\biggr]
\nonumber
\\[-8pt]
\\[-8pt]
\nonumber
&=& \arg\max_{1 \le k \le K}[ \mathbf{1'Z'}(\log\mathbf
{p}_k) + \log P(Y=k)].
\end{eqnarray}
This form of the naive Bayesian classifier is straightforward to
compute, especially in the software packages optimized for matrix
operations such as Matlab and R.

The naive Bayesian classifier \eqref{eqn:rule} can be easily extended
to accommodate differential losses to misclassification errors. Let
$W_{k',k}$ denote the cost of misclassifying a container to class $k'$
when it actually belongs to class $k$. Let $W_{k',k}=0$ when $k'=k$.
The expected loss of a classification is
$L(Y=k'|\mathbf{Z}) = \sum_{k=1}^K W_{k',k}P(Y=k|\mathbf{Z})$. The
optimal Bayesian decision rule is then
$Y_w = \arg\min_{k'} L(Y=k'|\mathbf{Z})$. Note that~\eqref{eqn:rule} is
a special instance using 0--1 loss of the decision rule $Y_w$.

The Bayesian classifier in this section assumes that the energy spectra
$\mathbf{p}_k$ is known for a variety of material classes. In the
remainder of this section we discuss the Bayesian learning to reduce
the uncertainty in estimating $\mathbf{p}_k$ for both lab and field
practices. Given $Y=k$, that is, the material class is known either by
experimental set up or by an offline detailed inspection, we consider
the conjugate prior for $\mathbf{p}_k$, that is, the $\operatorname{Dirichlet}(\theta
_{1,k},\ldots,\theta_{B,k})$ distribution. Then, training our classifier
on the data obtained from past containers with known classes, we
estimate $\mathbf{p}_k$ by
%
\begin{equation}\label{eqn:prior}
\hat p_{b,k} = \biggl(\sum_{i,d}R^i_{b,d}+\theta_{b,k}\biggr)\Big/\biggl(\sum
_{i,b,d}R^i_{b,d}+\sum_b\theta_{b,k}\biggr),
\end{equation}
where $R^i_{b,d}$ is the radiation count for the $i$th container in the
$k$th class. The summations over $i$ are over the containers with class
$k$ contents in the training data.

For determining the parameters of the conjugate priors, one would
typically use expert judgment. However, given the number of containers
passing through U.S. ports, a prior is unlikely to have any significant
effect on the ultimate decision except when the class is very rare.
Typically, conjugate priors are somewhat restrictive in terms of the
beliefs they can represent. \citet{Dalal1983} and \citet
{Diaconis1985} show that an appropriate mixture of natural conjugate
priors can approximate any arbitrary belief when the underlying
distributions belong to an exponential family. Further, the
corresponding posteriors converge almost surely to the true posterior.
Given these results, it is possible to extend this development to
mixtures of Dirichlet as priors for multinomial parameters. Using the
results shown in those papers, it can be shown that the resulting
estimates are the re-weighted mixtures of the estimates in \eqref{eqn:prior}.

\section{Asymptotic properties}\label{sec4}
We have so far shown that the proposed procedure is an optimal Bayesian classifier.
In this section we study the properties of our procedure for a fixed but arbitrary $K$,
by the asymptotic means assuming the Poisson model \eqref{eqn:model} and the multinomial
model \eqref{eqn:cond}, respectively.

We first consider the proposed classifier with $K$ classes with energy spectra
$\mathbf{p}_k, k=1,\ldots, K$, under the multinomial model \eqref{eqn:cond} by conditioning
on $N=\sum_d N_d$. For the development below, we consider the nontrivial case where all
energy spectra are positive and distinct corresponding to each of the material class.
Let $\mathbf{p}_{k'}$ be the true energy spectrum. The optimal Bayes classifier is $Y=k'$ if and only
if \eqref{eqn:rule} holds for $Y=k'$, namely,
\[
k' = \arg\max_k P(Y=k)P(\mathbf{Z}|Y=k).
\]
Taking log of ratio of the terms corresponding to $k'$ and $k$, this is true if and only if
\begin{equation}
\log \frac {P(\mathbf{Z}|Y=k') P(Y=k')} {P(\mathbf{Z}|Y=k) P(Y=k)} \ge 0 \qquad\mbox{for all }k.\label{eqn:8}
\end{equation}
Let $L_b=\sum_d R_{b,d}$ and $N=\sum_dN_d$. Since $(L_1,\ldots,L_B)|N$ follows a multinomial
distribution with parameters $(N;\mathbf{p}_{k'})$, by substituting the likelihood of
$P(\mathbf{Z}|Y)$ for $Y=k$ and $k'$,  we have \eqref{eqn:8} if and only if
\begin{equation}
 \frac{1}{N} \biggl[ \sum_b L_b \log \frac{p_{b,k'}}{p_{b,k}} + \log \frac{P(Y=k')}{P(Y=k)} \biggr] \ge 0.\label{eqn:18}
\end{equation}
Since $L_b$ follows binomial$(N, p_{b,k'})$, we have $|L_b/N - p_{b,k'}| = o_p(1)$. Further,
the second term in \eqref{eqn:18} is $O(N^{-1})$. Thus, the left-hand side of \eqref{eqn:18} is
%
\begin{eqnarray}\label{eqn:9}
&&\sum_b \biggl[p_{b,k'} \log \frac {p_{b,k'}}{p_{b,k}} +o_p(1) +O(N^{-1})\biggr] =
\eta(\mathbf{p}_{k'},\mathbf{p}_k)+ o_p(1)\\
\eqntext{\mbox{as } N \to \infty,}
\end{eqnarray}
where $\eta(\mathbf{p}_{k'},\mathbf{p}_k) = \sum_b p_{b,k'} \log \frac {p_{b,k'}}{p_{b,k}}$ is the
Kullback--Leibler divergence between the two multinomial distributions with parameters
$(1;\mathbf{p}_{k'})$, and $(1;\mathbf{p}_k)$. By the properties of the Kullback--Leibler
divergence and the Gibbs' inequality, $\eta(\mathbf{p}_{k'},\mathbf{p}_k)
\ge \inf_{s,s\ne k'} \eta(\mathbf{p}_{k'}, \mathbf{p}_s) >0$ for all $k \ne k'$.
Thus, $k'$ will be selected with probability approaching 1.

Now we generalize the above result, which says that our classifier is consistent for the
multinomial case to the general Poisson case without conditioning on $N$. For the Poisson
case, we have the total mean counts across $d$ as $\lambda_{k'} = \sum_{b,d} g(d,M)\lambda_{b,k'}$.
Suppressing the subscript $k'$, as $\lambda \to \infty$, $N/\lambda=1+ o_p(1)$, and thus, dividing
the left-hand side of equation~\eqref{eqn:8} by $\lambda$ again gives equation~\eqref{eqn:9}.
Thus, again, $k'$ will be selected with probability approaching 1. Summarizing this, we have the following theorem.

\begin{theorem}\label{thm1}
Under the multinomial model \eqref{eqn:cond} and the Poisson model \eqref{eqn:model}
and the assumption that all energy spectra $\mathbf{p}_1,\ldots, \mathbf{p}_K$ are
distinguishable, as $\lambda \to \infty$, for a given true class $k'$ in $1,\ldots, K$,
$P(Y=k'|\mathbf{Z}) \stackrel{p}{\rightarrow} 1$.
\end{theorem}

According to this result, irrespective of $K$, as long as all energy spectra are
distinct, as the counts become large, the classifier proposed here will converge
to the true underlying material.  Further, it follows from the proof that for given
counts, the probability of misclassification is higher for materials closer in the
Kullback--Leibler divergence sense.

We now consider robustness of our procedure to changes in the true underlying energy
distribution. By an argument that is an extension of the one used in proving the
above theorem, the following theorem can be shown similarly.
\begin{theorem}\label{thm2}
Under the multinomial model \eqref{eqn:cond} and the Poisson model \eqref{eqn:model},
and an arbitrary energy spectrum, $\mathbf{p}^*= (p_1^*,\ldots,p_B^*)$ not in
$\mathbf{p}_1,\ldots, \mathbf{p}_K$. Let $k'=\arg\min_{1\le k\le K} \eta(\mathbf{p}^*,\mathbf{p}_k)$.
Then as $\lambda \to \infty$, $P(Y=k'|\mathbf{Z}) \stackrel{p}{\rightarrow} 1$.
\end{theorem}

This result shows that even if the true distribution is not in the classification
scheme, the ones closest to it will be selected. In this sense we have robustness
with respect to variations in the underlying distribution.

\begin{remark*} Robustness with respect to correlated data.
In the development above, we assumed that $R_{b,d}$ are all independent over $d$.
One would expect by the underlying physics that radiation counts are independent
in different time intervals. Given that our development parallels the naive Bayes
models, it follows that the classifier \eqref{eqn:rule} is robust to this violation.
For further discussion of this we refer to \citet{Domingos1997} and \citet{Zhang2004}
who show robustness of the models with violations to independence.
\end{remark*}

In summary, the results indicate that our procedure is robust with respect to
variations in the underlying distribution and will scale up to large numbers
of categories as long as the counts are large. The counts increase if either
$M$ increases, the speed of driving through the portal decreases or the number
of $d$'s increase. Thus, from a policy perspective, for improving the detection
probability, the most attractive option is to pool across $d$'s.
It can also be shown by the Gibbs inequality that as the number of windows
increases by further refinement, the corresponding Kullback--Leibler divergence
between any two distributions also increases. Thus, for the next attractive
option, one needs to carry out a cost--benefit analysis between reducing the
speed and increasing the number of windows.
%
\begin{table}[b]
\caption{The energy spectra for the 6 pure material classes and
background. Numbers in parentheses are~$\lambda_{b,k}$}\label{table:p}
\begin{tabular*}{\textwidth}{@{\extracolsep{\fill}}ld{1.14}d{1.13}c@{}}
\hline
\textbf{Class} & \multicolumn{1}{c}{\textbf{Window 1}\phantom{.}}& \multicolumn{1}{c}{\textbf{Window 2}\phantom{0}} & \textbf{Window 3}\\
\hline
HEU & 0.954\mbox{ } (1.77\times10^4)& 0.033\mbox{ } (616) & 0.013 (247)\\
Fertilizer &0.635\mbox{ } (2.72\times10^3) &0.243\mbox{ } (1.0\times10^3) &0.122
(519)\\
Tile & 0.658\mbox{ } (2.22\times10^3)& 0.242\mbox{ } (818)&0.100 (338)\\
WGPu & 0.934\mbox{ } (6.09\times10^4)& 0.061\mbox{ } (3.9\times10^3)&0.005 (285) \\
Kitty litter &0.631\mbox{ } (1.7\times10^3)&0.292\mbox{ } (790)&0.077 (208)\\
Road salt & 0.662\mbox{ } (2.1\times10^3)&0.273\mbox{ } (873)&0.065 (208)\\
Background &0.651\mbox{ } (1.4\times10^3)&0.249\mbox{ } (519)&0.100 (207)\\
\hline
\end{tabular*}
\end{table}

\eject
\section{Numerical studies}\label{sec5}
\subsection{Classification of all classes}\label{sec5.1}
To explore and illustrate the sensitivity and specificity of our
method, we consider a number of simulated examples and scenarios. The
first example is based on energy spectra emitted by the following 7
classes of material: WGPu, HEU, fertilizer, tiles (refractory
material), kitty litter, road salt and background (i.e., a container's
radiation is undetectable from the background). The energy spectra were
synthesized from a few sources, including \citet{Ely2004} and \citet
{Ely2006}, presentations from Pacific Northwest National Laboratory
(PNNL) and consultation with PNNL scientists. For these data, the
reported energy spectra only consisted of 3 windows. Table \ref
{table:p} lists the energy spectra of mean counts for 1 unit of the 6
pure material classes and background at the distance defined as $d=0$.
The man-made nuclear material classes were chosen based on their
importance for detection in ports,
while the NORM classes were selected based on their probability of
misclassification indicated in Table \ref{table:fp} (courtesy of PNNL).
Radiation from the NORM classes is primarily from potassium-40, which
naturally exists in many common materials.
For reference, false positive probabilities for some typical NORM
materials are shown in Table \ref{table:fp} (compiled from personal
communications with PNNL), per the currently deployed method and
publicly available data.
The units of quantity are different for different classes. \citet
{Ely2004} and \citet{Ely2006} considered 1 unit WGPu as 99.4 g in a
powdered oxide form and doubly contained in schedule-80 stainless steel
closed pipes that provide shielding. Even with shielding, 1 unit WGPu
is still highly radioactive. 1 unit HEU was considered as 123 g of
93.1\% enriched uranium consisting of a number of stacked foils, which
is a moderately strong radiation source. For all NORM material classes,
1 unit is 5 kg. We also consider another man-made class as a mixture of
0.5 unit HEU and 0.5 unit of WGPu.

%
\begin{table}
\caption{Some frequently misclassified NORM classes by the currently
deployed methodology. Numbers are proportions of false positive for
detecting radiation risk}
\label{table:fp}
\begin{tabular*}{\textwidth}{@{\extracolsep{\fill}}lccc@{}}
\hline
\textbf{Source class}& \textbf{Port A} &\textbf{Port B}& \textbf{Port C}\\
\hline
Kitty litter & 0.34 & 0.25 & -- \\
Abrasive pads & 0.14 & 0.05 & -- \\
Mica & 0.05 &-- &--\\
Fertilizer & 0.05 & 0.13 & --\\
Ceramics/tile & 0.04 & 0.09 & 0.28 \\
Granite & 0.04 &-- & 0.10\\
Salt & -- & 0.05 & --\\
Trucks/cars & 0.02& -- & --\\
Aluminum & -- & 0.15 & --\\
Other metal & -- & 0.03 & -- \\
\hline
\end{tabular*}
\end{table}

To numerically simulate the radiation data, we set up $g(d,M)=(1-d)M$
and 20 distances $(0.9, 0.8, \ldots, 0.1, 0, 0, 0.1, \ldots, 0.9)$,
corresponding to the portal passing-through time for a container. As
discussed previously, the specific form of $g(d,M)$ does not affect the
Bayesian classifier, but is only needed for generating data. This
simulation scenario is set up to have the total counts $\sum_d N_d$
comparable to the actual scanning process shown in Figure \ref{fig:nex}.
In the first simulation study, we generated 10,000 samples of Poisson
variables at each $d$ for a variety of material classes. We assign a
prior probability $10^{-9}$ to each man-made source, and use equal
prior probability for each NORM class.
Table \ref{table:single} reports the misclassification probabilities.
The Bayesian classifier has excellent performance in this scenario.
It was able to classify all materials correctly except for minor
confusion between tiles and background, both not dangerous.
In particular, there is no misclassification between man-made source
and NORM in 10,000 simulations.

%
\begin{table}
\caption{Summary of classification simulations with 8 classes. Classes
A to H are WGPu (A), HEU (B), mixture of 0.5 unit. WGPu and 0.5 unit
HEU (C), fertilizer (D), tile (E), kitty litter (F), salt (G) and background (H)
}\label{table:single}
\begin{tabular*}{\textwidth}{@{\extracolsep{\fill}}lccccd{1.3}ccd{1.3}@{}}
\hline
& \multicolumn{8}{c@{}}{\textbf{Classified}}\\[-6pt]
& \multicolumn{8}{c@{}}{\hrulefill}\\
\textbf{True class} &\textbf{A}&\textbf{B}&\textbf{C}&\textbf{D}&\multicolumn{1}{c}{\textbf{E}}&\textbf{F}&\textbf{G}&\multicolumn{1}{c@{}}{\textbf{H}}\\
\hline
A&1&0&0&0&0&0&0&0\\
B&0&1&0&0&0&0&0&0\\
C&0&0&1&0&0&0&0&0\\
D&0&0&0&1&0&0&0&0\\
E&0&0&0&0&0.937&0&0&0.063\\
F&0&0&0&0&0&1&0&0\\
G&0&0&0&0&0&0&1&0\\
H&0&0&0&0&0.111&0&0&0.889\\
\hline
\end{tabular*}
\end{table}

\begin{remark*} At this scale of total counts, the
classifier is relatively insensitive to the choice of prior $P(Y=k)$.
From \eqref{eqn:finalform}, it can be seen that as long as $\log
(P(Y=k)) = o_p(\mathbf{1'Z})$, the prior should have no practical
impact on the classifier. Recall that the actual counts are large (see
Figure \ref{fig:nex}). Hence, the prior probability for man-made source
is unlikely to influence the classification remarkably. Since the total
counts in simulations are also large (comparable to Figure \ref
{fig:nex}), there is not a single change in misclassifying a man-made
source to NORM, or vice versa, if we use the equal prior probability on
all classes, or change the prior probability of each man-made source to
$10^{-9}$.
\end{remark*}

We now study a more complex situation by considering a more devious
terrorist strategy of mixing materials.
The man-made source possessed by terrorists, such as WGPu or HEU, may
be mixed with a NORM material class. Moreover, the total count may be
made small enough to pass the current inline inspection.
Hence, to challenge the classifier, the main sensitive material classes
should allow mixtures of man-made and NORM classes.
For hard to detect situations, the mixture should have a small quantity
of a man-made source and a relatively large quantity of NORM.
In the next numerical study, we consider this possibility by using a
list of mixtures of the known classes except for background. Each
mixture class mixes WGPu or HEU with one of the 4 NORM classes.
Assuming that the mixture is noninterfering, the energy spectrum for
the mixture of two classes $k$ and $k'$ with quantities $M$ and $M'$
will be given by
%
\begin{equation}
\frac{ [(1-d)M\lambda_{b,k}+(1-d)M'\lambda_{b,k'}]}{\sum
_b[(1-d)M\lambda_{b,k}+(1-d)M'\lambda_{b,k'})]},\qquad  b=1,\ldots, B.
\end{equation}

\begin{figure}[b]

\includegraphics{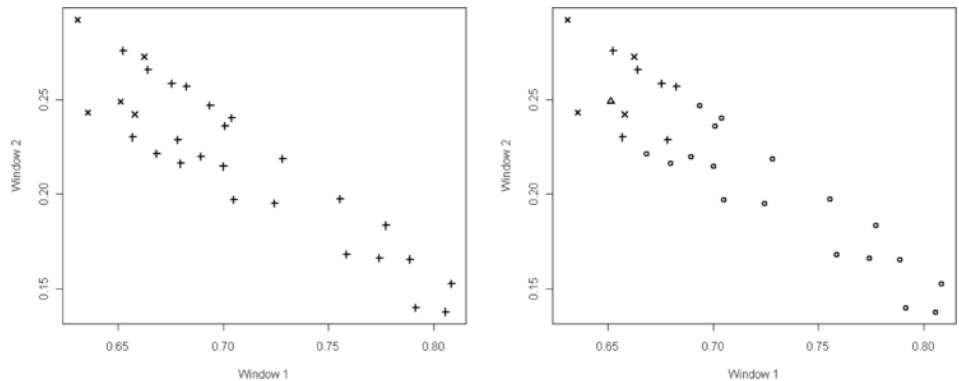}

\caption{Illustrations of classes. The left frame has all 29 classes as
benchmark and the right frame has 10 automatic selected benchmarks.
Circle: dangerous nonbenchmark; cross: dangerous benchmark; $\times$:
nondangerous benchmark; triangle: nondangerous nonbenchmark.}\label{fig:c}
\end{figure}

We considered 3 small quantities for each man-made source class to
simulate the different diluting effects.
The quantity of HEU was 0.025, 0.05 or 0.1 unit, and the quantity of WGPu
was 0.005, 0.01 or 0.025 unit. The quantity of HEU is slightly larger than
WGPu due to the relatively weaker radioactivity by the definition of 1
unit HEU.
All of the classes being inspected except for background have the same
mean count in the first window. Since the first window has much larger
counts than the other two windows, a real container of all these
classes will have approximately the same total counts.
We set the mean count at distance $d=0$ in the first window, $M\lambda
_{1,k}+M'\lambda_{1,k'}$ at $3000$ for all mixture classes.
The quantity of all NORM classes is solved from the preceding
restriction. For example, the mixture class ``0.025HEU${}+{}$Fertilizer'' has
0.025 unit HEU and 0.941 unit fertilizer.
The mean value $3000$ was set to be close to 1 unit of NORM and far
less than~1 unit of a man-made source,
representing a scenario that could be encountered at port inspection.
It is likely that the currently deployed method will not detect most of
the containers with small quantities of HEU or WGPu, since it is based
on maximal gross counts.
We also considered the 4 pure NORM classes and the background. Except
for background, the quantities of the 4 NORM classes were set according
to $M\lambda_{1,k}=3000$.
For instance, the pure kitty litter class under this constraint has
1.759 units. This gave a total of 29 material classes to be classified.
We call the mixture classes with man-made source as \textit{dangerous}
classes, and the NORM classes and background as \textit{nondangerous}
classes. Since the material classes have energy spectra consisting of 3
windows, we can plot the corresponding proportion of spectra in two
energy windows, which is shown in Figure \ref{fig:c}. As can be seen,
some dangerous and nondangerous classes are intermingled.

The Bayesian classifier again has very good performance in this
scenario. For all dangerous classes, the misclassification probability
is below 0.005 (including misclassified as other dangerous classes).
Most of the misclassification probabilities for NORM classes are below
0.01, except for the misclassifications between tile and background (0.06
for misclassifying tile as background, and 0.11 for misclassifying
background as tile). This indicates that the Bayesian classifier scales
well and is hard to defeat by the simple strategy of mixing a man-made
nuclear material with a NORM.

\subsection{Classification into dangerous versus nondangerous classes}\label{sec5.2}
In the previous section we examined the problem of exact classification
of all materials, and, consequently, we used all material classes to
build the classifier. Here we consider a more limited objective for
preliminary screening, namely, classify a given material as dangerous
or nondangerous from a given large list of $K$ dangerous (material
involving man-made nuclear sources including mixtures) and nondangerous
materials. After classifying in these categories, one can use a scheme
akin to the previous section to identify the specific dangerous
material at the next stage.

Given this limited objective, and that there may not be any natural
differentiation (e.g., hyperplane in some transformed space) in the
spectra profile of dangerous and nondangerous materials, this problem
cannot be solved by the standard binary classification methods.
Instead, we consider the following scheme, whereby we build the
classifier based on a subset of materials (called \textit{benchmark
classes}) from the list. A material is classified as dangerous if the
corresponding selected benchmark material is dangerous; otherwise it is
nondangerous.

For this setting, the simplest scheme would be to use all materials as
benchmark classes. The question we address is the following: can one be
more parsimonious in the number of benchmark materials to build the
classifier? Having a smaller number of benchmark classes of materials
should help in scaling up the solution. Below we first investigate this
question in the context of 29 materials considered in the last
subsection and compare this ``all classes'' solution with an
algorithmically derived solution.

The algorithm to select benchmark classes is iterative and does not
require specific structure of the materials. It is
motivated by the support vector machine method and forward selection
method [\citet{Hastie2008}]. However, unlike the support vector machine
method, which finds the separating hyperplane with the biggest margin,
we identify high leverage points that are difficult to distinguish and
use them as benchmark classes and we iterate on the scheme by forward selection.
Given that the total counts are large in our applications, for
identifying a high leverage point, we revert to Theorem \ref{thm2}, which states
that if a class is not in the classifier, the class nearest in the
Kullback--Leibler sense will be selected by the classifier. For
describing the algorithm, the following notation is in order. Suppose
that of the $K$ major distinct classes, $K_1$ classes are dangerous,
and $K_2$ classes are nondangerous, all with distinct energy spectra.
Let $D$ and $ND$ be the corresponding sets of material. At the $i$th
iteration, the algorithm produces a list of dangerous and nondangerous
material to be used as the benchmark class for the next stage, and let
us denote those subclasses as $D_i$ and $ND_i$. Finally, let $\eta
(\mathbf{p},A)=\inf\{\eta(\mathbf{p},\mathbf{p}_k), k \in A\}$, where
$\eta(\mathbf{p},\mathbf{p}_k)$ is the Kullback--Leibler divergence
defined in Section \ref{sec4}. The algorithm for selecting benchmark classes is
as follows.

\begin{algorithm*}\vspace*{-6pt}
\begin{longlist}
\item[(a)] For $i=1$, initiate by taking $D_1$ and $ND_1$ to be any pair of
dangerous and nondangerous material which are nearest in $\eta^*$, the
symmetrized version of $\eta$ [i.e., $\eta^*(\mathbf{p},\mathbf{q})=\eta
(\mathbf{p},\mathbf{q})+\eta(\mathbf{q},\mathbf{p})$].
\item[(b)] At the stage $i+1$, let $AD_{i+1}= \{k\dvtx k \in D - D_i,   \ \eta
(\mathbf{p}_k,D_i) - \eta(\mathbf{p}_k,{ND_i}) > 0\}$ and $AND_{i+1}= \{
k\dvtx k\in ND-ND_i,   \eta(\mathbf{p}_k,ND_i)- \eta(\mathbf{p}_k,D_i) >
0\}$. Note that if the total counts are large, $AD_i$ will be
misclassified with high probability as nondangerous and vice versa.
\item[(c)] If $AD_{i+1}$ and $AND_{i+1}$ are empty, then stop. Otherwise
$D_{i+1} = D_i \cup k^*$, where
\[
k^* = \arg\min_{k \in AD_{i+1}} \eta(\mathbf{p}_k,ND_i).
\]
Similarly construct $ND_{i+1}$.
\item[(d)] Repeat this process till $AD_{i+1}$ and $AND_{i+1}$ are empty.
Note that when we stop, we have asymptotically probability 1 of correct
classification of materials in dangerous and nondangerous classes.
\end{longlist}
\end{algorithm*}

In our previous example with 29 classes, the above algorithm leads to
only 10 benchmark classes.
Table \ref{table:s2} gives the corresponding probabilities of misclassification.
All solutions have exceptional performance and are indistinguishable.
Based on this comparison, if the binary classification is the main
objective, it would be always preferable to use the algorithmic
approach, since it reduces the complexity without penalizing
performance. Further, it does not require any specific structure of the
materials.

%
\begin{table}[b]
\caption{Summary of simulations with 29 classes: numbers are the
probability of detecting dangerous radioactive materials. The two
columns under each class correspond to all 29 classes as benchmark and
the automatic selected benchmark. The selected benchmark classes are
marked with a star}\label{table:s2}
\begin{tabular*}{\textwidth}{@{\extracolsep{\fill}}llllll@{}}
\hline
0.025HEU${}+{}$Fertilizer&1&1& 0.005WGPu${}+{}$Fertilizer*&1&1\\
0.05HEU${}+{}$Fertilizer&1&1 & 0.01WGPu${}+{}$Fertilizer&1&1\\
0.1HEU${}+{}$Fertilizer&1&1& 0.025WGPu${}+{}$Fertilizer&1&1\\
0.025HEU${}+{}$Tile&1&1& 0.005WGPu${}+{}$Tile*&1&1\\
0.05HEU${}+{}$Tile&1&1 & 0.01WGPu${}+{}$Tile&1&1\\
0.1HEU${}+{}$Tile&1&1 & 0.025WGPu${}+{}$Tile&1&1\\
0.025HEU${}+{}$Kitty litter*&0.997&0.997&0.005WGPu${}+{}$Kitty litter*&0.999&1\\
0.05HEU${}+{}$Kitty litter&1&1&0.01WGPu${}+{}$Kitty litter*&0.999&1\\
0.1HEU${}+{}$Kitty litter&1&1&0.025WGPu${}+{}$Kitty litter&1&1\\
0.025HEU${}+{}$Salt&1&1&0.005WGPu${}+{}$Salt*&1&1\\
0.05HEU${}+{}$Salt&1&1&0.01WGPu${}+{}$Salt&1&1\\
0.1HEU${}+{}$Salt&1&1&0.025WGPu${}+{}$Salt&1&1\\
Fertilizer*&0&0&\multicolumn{1}{l}{Tile*}&0&0\\
Kitty litter*&0&0&\multicolumn{1}{l}{Salt*}&0.011&0.012\\
Background&0&0\\
\hline
\end{tabular*}
\end{table}

Next, we examine the performance of the algorithm in a large study with
100 classes, consisting of 75 dangerous and 25 nondangerous.
The dangerous classes were mixtures of a man-made source with 4 NORM
classes. The quantity of man-made source was either 0.025 unit HEU or
0.005 WGPu, and the quantities of 4 NORM classes were generated from a
Dirichlet distribution with parameters $(0.25,0.25,0.25,0.25)$.
The nondangerous classes were mixtures of 4 NORM classes, and the
quantities were generated independently from $U(0,0.25)$.
This scenario is more stringent than the previous scenarios, since this
scenario uses the smallest quantities of man-made sources used in the
previous scenarios.
We generated 50 sets of classes. The mean number of benchmark classes
chosen by the algorithm is 6.9 and the standard deviation (sd) is 5.7.
The median number of benchmark classes is 5 and the median absolute
deviation (mad) is 6.
For each of the 50 sets of generated classes, we ran 1000 simulations
to evaluate the performance. The mean probability of misclassifying a
dangerous class as nondangerous is 0.001 (median${}={}$0.000, sd${}={}$0.009,
mad${}={}$0.000), and the mean probability of misclassifying a nondangerous
class as dangerous is 0.008 (median${}={}$0.000, sd${}={}$0.037, mad${}={}$0.000).

In summary, the numerical examples in this section show that the
parsimonious benchmark classes chosen by the proposed algorithm retain
the capability of detecting illicit nuclear material and are more
efficient in computation.

\section{Discussion}\label{sec6}
We have proposed a Bayesian approach for modeling the energy
distribution as well as the total energy emitted by an unknown material class.
We have also proposed a Bayesian decision rule for classifying a new
container into one of the known classes.
Our approach uses all available data compared to the currently deployed
method, which only uses the maximal counts.
We have examined its robustness properties by simulations and
asymptotic arguments and have shown that the proposed approach is
scalable. For binary classification between dangerous and nondangerous
materials, we have proposed a scalable algorithm for selecting a small
number of benchmark classes motivated by the support vector machine and
forward selection methods. The results in this paper are encouraging
compared to the false positives reported by the currently deployed
method in Table~\ref{table:fp}.

Since our approach is based on classification, two questions naturally
emerge for our and other classification approaches, namely, (1) to what
extent is the procedure robust with respect to variation in the
underlying distribution? and (2) to what extent can one entertain the
possibility of none of the above classes? Theorem \ref{thm2} and ensuing
discussion show that our approach is robust in the sense that even if
the true distribution is not in the classifier, the benchmark class
closest to it will be selected. While our and all other classification
approaches do not allow a direct answer to the second question, it
should be feasible to perform a Chi-squared goodness of fit test for
the selected class. If such a test rejects the hypothesis, then one
possibility is to screen such a container. Clearly, the efficacy of
such a simultaneous procedure needs to be further investigated.
Also, since our approach does not depend upon the number of windows and
the time to pass-through the portal,
it is likely to be easily extendable to newer portals, more windows,
material classes and changes in design.

\section*{Acknowledgments} The authors wish to thank James Ely, Dennis
Weier, Rick Bates and the rest of the PNNL team for sharing information
and helpful discussions, and Fred Roberts and Tami Carpenter of DIMACS
at Rutgers University for getting the authors involved in this problem
of national security. The authors also thank Colin Mallows of Avaya
Labs for helpful
comments, and the referee and Associate Editor for helpful comments.

\printaddresses

\end{document}